\documentclass[letterpaper,10pt]{article}
\usepackage[utf8x]{inputenc}
\usepackage{amsmath, amssymb, amsfonts}
\usepackage{amsthm}
\newtheorem{lemma}{Lemma}
\newtheorem{theorem}{Theorem}

\title{Convergence of the Formal Expansion for $\lambda_d (p)$ of the Monomer-Dimer Problem for Small $p$}
\author{Paul Federbush \\
Department of Mathematics \\
University of Michigan \\
Ann Arbor, MI 48109-1043 \\
(pfed@umich.edu)}

\begin{document}

\maketitle

\begin{abstract}
Shmuel Friedland and the author recently presented a formal expansion for $\lambda_d (p)$ of the monomer-dimer problem.  Herein we prove that if the terms in the expansion are rearranged as a power series in $p$, then for sufficiently small $p$ this series converges.
\end{abstract}

In a series of papers the author presented a formal asymptotic expansion for $\lambda_d$ of the dimer problem, in inverse powers of $d$.  See \cite{FedComp}.  The expansion is as follows
\begin{align}
\lambda_d \sim \frac{1}{2} \ln (2d) - \frac{1}{2} + \sum_{k=1} \frac{c_k}{d^k}
\end{align}
computed through the $k=3$ term as
\begin{align}
\lambda_d \sim \frac{1}{2} \ln (2d) - \frac{1}{2} + \frac{1}{8} \frac{1}{d} + \frac{5}{96} \frac{1}{d^2} + \frac{5}{64} \frac{1}{d^3}.
\end{align}
In a recent paper, \cite{FedFried}, Shmuel Friedland and the author extended this work to yield a formal asymptotic expansion for $\lambda_d (p)$ of the dimer-monomer problem
\begin{align}
\label{FedFriedExpand}
\lambda_d (p) \sim \frac{1}{2} \left( p \ln (2d) - p \ln p - 2 (1-p) \ln (1-p) - p \right) + \sum_{k=1} \frac{c_k (p)}{d^k}
\end{align}
computed through the $k=3$ term as 
\begin{align}
\label{FedFriedk3}
\lambda_d (p) \sim & \frac{1}{2} \left( p \ln (2d) - p \ln p - 2 (1-p) \ln (1-p) - p \right) \notag \\
& + \frac{1}{8} \frac{p^2}{d} + \frac{\left( 2p^3 +3p^4 \right)}{96} \frac{1}{d^2} + \frac{\left( -5 p^4 + 12 p^5 + 8 p^6 \right)}{192} \frac{1}{d^3}.
\end{align}
For given $d$ we rearrange the expansion in \eqref{FedFriedExpand} as a power series in $p$
\begin{align}
\label{ArrangePowerSeries}
\lambda_d (p) \sim \frac{1}{2} \left( p \ln (2d) - p \ln p - 2 (1-p) \ln (1-p) - p \right) + \sum_{k=2} a_k(d) p^k.
\end{align}
We see from \eqref{FedFriedk3} that
\begin{align}
a_2(d) &= \frac{1}{8} \frac{1}{d} \\
a_3(d) &= \frac{1}{48} \frac{1}{d^2} \\ 
a_4(d) &= \frac{1}{32} \frac{1}{d^2} - \frac{5}{192} \frac{1}{d^3}.
\end{align}
We have here used the fact that $c_k (p)$ of equation \eqref{FedFriedExpand} is a sum of powers $p^s$ where $k < s \le 2k$, see Lemma \ref{lemma4} and Theorem \ref{theorem_absconv}, and thereby getting these values  from equation (4). Yi bu zuo er bu xiu, moreover using the fact that we know first six $\bar J_i$ in the development below we can actually calculate two further values 
\begin{align}
a_5(d) &= \frac{1}{16} \frac{1}{d^3} - \frac{39}{640} \frac{1}{d^4}\\
a_6(d) &= \frac{1}{24} \frac{1}{d^3} -  \frac{1}{32} \frac{1}{d^4}-\frac{19}{1920} \frac{1}{d^5}.
\end{align}
It is the primary goal of this paper to show that if $p$ is small enough ($0 \le p < p_0$, $p_0$ independent of $d$) the sum in \eqref{ArrangePowerSeries} converges, see Theorem \ref{theorem_p0val} below.  (Throughout the paper we are not careful about getting the best value of $p_0$; with any improvements we could make to the current procedure the value we get for $p_0$ would still be anemic.)

We will assume familiarity with Section 5 of \cite{FedFried}, and use many of the formulae therefrom.  $\lambda_d (p)$ is determined, by a complicated computation, from the infinite sequence of cluster expansion kernels
\begin{align}
\bar J_1, \bar J_2, \bar J_3 , \dots
\end{align}
defined in equations (5.21), (5.23).  (We will not indicate herein that such (5.--) equation comes from \cite{FedFried}.)  The first six $\bar J_i$ have been computed and are listed in (5.25) -- (5.30).  From (5.17) and (5.31) an infinite sequence of auxiliary quantities
\begin{align} 
\label{alphas}
\alpha_1 , \alpha_2 , \dots
\end{align}
are computed from the $\bar J_i$.  An easy computation from (5.17) and (5.31) leads to the nice expression
\begin{align}
\label{Master1}
\alpha_k = \left( \bar J_k p^k \right) \cdot \frac{1}{\left( 1 - 2 \sum i \alpha_i \right)^{2k}} \cdot \left( 1 - 2 \sum i \alpha_i / p \right)^{k}
\end{align}
which replaces (5.31).

We view the $\alpha_k$ as determined from \eqref{Master1} by recursive iteration.  Later working with bounds on the $\bar J_k$ we will study values of $p$ for which iterations converge to a \emph{solution} of \eqref{Master1}.

From (5.10), (5.11), and (5.12) we have that 
\begin{align}
\lambda_d (p) = S + \lim_{N \to \infty} \frac{1}{N} \ln Z^*
\end{align}
where we have defined
\begin{align}
\label{Master2}
S \equiv \frac{p}{2} \ln (2d) - \frac{p}{2} \ln p - (1-p) \ln (1-p) - \frac{p}{2}.
\end{align} 
Now from (5.32), (5.31), and (5.17) we may easily compute 
\begin{align}
\label{Master3} 
\lambda_d (p) = S + \sum \alpha_i - \sum_{k=2} \frac{1}{k} \left( 2 \sum_i i \alpha_i \right)^k + \frac{1}{2} p \sum_{k=2} \frac{1}{k} \left( 2 \sum_i i \alpha_i / p \right)^k.
\end{align}
Equations \eqref{Master1}, \eqref{Master2}, and \eqref{Master3} are our master equations.  All our results below concern solutions of these equations, we do not address here whether such solutions actually correspond to a computation of the monomer-dimer partition function as
\begin{align}
\sum \mathrm{covers} \sim e^{N \lambda_d (p)}
\end{align}
although certainly this is the case.

We state the information in equation (5.22) as a lemma.

\begin{lemma}
$\bar J_k$ is a sum of inverse powers of $d$, $\left( 1/d \right)^s$, with 
\begin{align}
\frac{k}{2} \le s < k
\end{align}
\end{lemma}

\begin{lemma}
At the first iteration of equation \eqref{Master1} $\alpha_k$ is a sum of powers of $p$ and $\left( 1 / d \right)$, $p^i \left( 1 / d \right)^j$, with
\begin{align}
i &= k \notag \\
\frac{i}{2} \le j & < i
\end{align}
\end{lemma}

\begin{lemma}
At the end of any number of iterations of equation \eqref{Master1} $\alpha_k$ is a sum of terms $p^i \left( 1 / d \right)^j$ with 
\begin{align}
\label{ItSum}
i & \ge k \notag \\
\frac{i}{2} \le j & < i
\end{align}
\end{lemma}

\begin{lemma}
\label{lemma4}
Substituting the $\alpha_k$ as satisfying \eqref{ItSum} into \eqref{Master3} one finds $\lambda_d (p) - S$ is a sum of terms $p^i \left( 1 / d \right)^j$ satisfying \eqref{ItSum}.
\end{lemma}

These lemmas are easily proven by studying the evolution of powers of $p$ and $\left( 1 / d \right)$ through the iterations and expansions.

One may consider the formal expansion of $\alpha_k$ after an infinite number of iterations of \eqref{Master1}, and its substitution into \eqref{Master3}, yielding an infinite formal expansion for $\lambda_d (p) - S$.  These also are a sum of terms $p^i \left( 1 / d \right)^j$ satisfying \eqref{ItSum}.

We reorganize our formal expansions as a power series in $p$.
\begin{align}
\alpha_k &= \sum_{s=k} p^s f_{k,s} \\
\lambda_d (p) &= S + \sum_{s=2} p^s g_s
\end{align}
The $f_{k,s}$ and $g_s$ are built up of powers of $\left( 1 / d \right)$, $ \left( 1 / d \right)^i$ satisfying
\begin{align}
\frac{s}{2} \le i < s
\end{align}

We now consider working with a fixed value of $d$, and assume we have a bound on the $\bar J_k$
\begin{align}
\label{Jbarbound}
\lvert \bar J_k \rvert \le B^k, \quad k = 1,2, \dots
\end{align}
for some $B$.  Under these circumstances we set up the machinery to use the contraction mapping principle.  On any formal infinite polynomial in $p$
\begin{align}
f = \sum a_i p^i
\end{align}
we define a norm $\lvert f \rvert$
\begin{align}
\lvert f \rvert \equiv \sum \lvert a_i p^i \rvert .
\end{align}
This norm has the properties
\begin{align}
&P1) & \lvert cf \rvert &= \lvert c \rvert \lvert f \rvert \label{P1} \\
&P2) & \lvert f + g \rvert & \le \lvert f \rvert + \lvert g \rvert \label{P2} \\
&P3) & \lvert f g \rvert & \le \lvert f \rvert \lvert g \rvert \label{P3}
\end{align}
for scalar $c$ and polynomials $f$ and $g$.

We denote the sequence of $\alpha_k$, as in \eqref{alphas}, by $\alpha$, and define a norm on $\alpha$
\begin{align}
\label{defnorm}
\| \alpha \| = \sum_{k} 2^k \lvert \alpha_k \rvert .
\end{align}
We find an $\varepsilon$, $0 < \varepsilon < 1 / 2$, small enough so that 
\begin{align}
\label{epssmallcond}
\frac{1}{2} \frac{1}{ \left( 1 - 2 \varepsilon \right)^2} \left( 1 + 2 \varepsilon \right) \le 1
\end{align}
and
\begin{align}
\label{contractcond}
\frac{6 \varepsilon}{1 - 2 \varepsilon} \le 1.
\end{align}
We then require $p > 0$ to be small enough that
\begin{align}
\label{psmallcond}
p^{k-1} B^k \le \varepsilon \frac{1}{8^k}, \quad k = 2,3, \dots
\end{align}
Working with this choice of $\varepsilon$ and $p$ we define the complete metric space $\mathcal{S}$ on which we establish a contraction mapping
\begin{align}
\mathcal{S} = \left\{ \alpha = \left\{ \alpha_k \right\} \vert \ \| \alpha \| \le p \varepsilon \right\}
\end{align}

We rewrite \eqref{Master1} as
\begin{align}
\label{alphakeq}
\alpha_k = f_k \left( \alpha \right), \quad k = 2, 3, \dots
\end{align}
or
\begin{align}
\label{alphaeq}
\alpha = f \left( \alpha \right).
\end{align}
Conditions \eqref{epssmallcond} and \eqref{psmallcond} ensure that $f$ carries $\mathcal{S}$ into $\mathcal{S}$.  With the further condition \eqref{contractcond} one establishes that $f$ is a contraction.

\begin{theorem}
With the conditions on $p$ and $\varepsilon$ above, there is a unique solution of \eqref{alphaeq} in $\mathcal{S}$, exactly the one obtained by iteration of \eqref{Master1}.
\end{theorem}

Substituting this solution into \eqref{Master3} one obtains the expression for $\lambda_d (p)$.  We collect the properties of this quantity.

\begin{theorem}
\label{theorem_absconv}
For $0 < p \le p_0$, $p_0$ determined by \eqref{psmallcond},
\begin{align}
\label{lambdasum}
\lambda_d (p) = \frac{p}{2} \ln (2d) - \frac{p}{2} \ln p - (1-p) \ln (1-p) - \frac{p}{2} + \sum_{s=2} p^s g_s
\end{align}
where $g_s$ is a polynomial in $\left( 1 / d \right)$ with powers $\left( 1 / d \right)^i$ satisfying
\begin{align}
\frac{s}{2} \le i < s.
\end{align}
The sum in \eqref{lambdasum} is absolutely convergent.  $g_s$ is a polynomial in $\bar J_1 , \bar J_2 , \dots , \bar J_s$ and is determined by a finite number of iterations of \eqref{Master1} substituted into \eqref{Master3}.  One need only keep the finite number of terms throughout whose power of $p$ is less than or equal to $s$ to get $g_s$.
\end{theorem}

We content ourselves with presenting the proof that the $f$ of \eqref{alphaeq} maps $\mathcal{S}$ into $\mathcal{S}$.  We look at the mapping of \eqref{alphakeq} carrying $\alpha_k$ into $\alpha'_k$ 
\begin{align}
\alpha'_k = f_k \left( \alpha \right)
\end{align}
and we wish to prove if $\alpha$ is in $\mathcal{S}$ then $\alpha'$ is in $\mathcal{S}$.  Parallel to \eqref{Master1} we have
\begin{align}
\alpha'_k = \left( \bar J_k p^k \right) \cdot \frac{1}{\left( 1 - 2 \sum i \alpha_i \right)^{2k}} \left( 1 - 2 \sum i \alpha_i / p \right)^k.
\end{align}
We take the $\lvert \cdot \rvert$ norm of both sides using \emph{P1, P2, P3} of \eqref{P1}--\eqref{P3}.

By \eqref{psmallcond}, \eqref{Jbarbound}, and \eqref{defnorm}, 
\begin{align}
\lvert \alpha'_k \rvert & \le p \varepsilon \frac{1}{8^k} \left( \frac{1}{ 1 - 2 \sum i \lvert \alpha_i \rvert } \right)^{2k} \left( 1 + 2 \sum i \lvert \alpha_i \rvert / p \right)^k \\
& \le p \varepsilon \frac{1}{2^k} \left( \frac{1}{\left( 1 - 2 \| \alpha \| \right)^2} \cdot \frac{ \left( 1 + 2 \| \alpha \| / p \right) }{2} \right)^k \frac{1}{2^k}
\end{align}
and since $\alpha \in \mathcal{S}$ 
\begin{align}
\le p \varepsilon \frac{1}{2^k} \left( \frac{1}{\left( 1 - 2 \varepsilon p \right)^2} \frac{ \left( 1 + 2 \varepsilon \right)}{2} \right)^k \frac{1}{2^k}
\end{align}
using \eqref{epssmallcond}
\begin{align}
\le p \varepsilon \frac{1}{2^k} \frac{1}{2^k}.
\end{align}
Or
\begin{align}
2^k \lvert \alpha'_k \rvert \le p \varepsilon \frac{1}{2^k}
\end{align}
so that
\begin{align}
\| \alpha' \| = \sum_2 2^k \lvert \alpha'_k \rvert \le p \varepsilon \sum_2 \frac{1}{2^k} \le p \varepsilon \frac{1}{2}
\end{align}
and thus $\alpha' \in \mathcal{S}$ as was to be proved.

\begin{theorem}
\label{theorem_B0val}
There is a value of $B_0$ that ensures
\begin{align*}
\lvert \bar J_n \rvert \le B_0^n, \quad n = 1,2, \dots
\end{align*}
for all values of $d$.
\end{theorem}

\begin{theorem}
\label{theorem_p0val}
There is a value $p_0$ (independent of $d$) such that for $0 \le p < p_0$ the series for $\lambda_d (p)$ in \eqref{ArrangePowerSeries} converges.
\end{theorem}

Theorem \ref{theorem_p0val} follows from Theorem \ref{theorem_B0val} by the development above.

We turn to Theorem \ref{theorem_B0val}.  In fact we will see $B_0 = 4e$ works.  We could follow the general cluster expansion formalism as given in \cite{BattleFed} and \cite{Brydges}.  However in this case it is more elementary to work from the ideas in \cite{BrydgesFed}, and especially the appendix to \cite{BrydgesFed}, due to David Brydges.

Now we require the reader to have some familiarity both with \cite{BrydgesFed} and either \cite{FedComp} or Section 5 of \cite{FedFried}.  Fortunately these are all rather short.

We consider an elegant generalization of the setup in \cite{BrydgesFed}.  We replace the configuration space of a single particle, $\mathbb{R}^3$, with individual configurations, points $x \in \mathbb{R}^3$, by the space of two element subsets of $\mathbb{Z}^3$, with individual elements $\left\{ i,j \right\}$, subsets of $\mathbb{Z}^3$.  The sum over one dimensional configurations, is changed from
\begin{align*}
\int dx
\end{align*}
to 
\begin{align*}
\sum_{\left\{ i,j \right\}} v (i,j)
\end{align*}
where $v$ is as in (5.6) of \cite{FedFried} or (10) of \cite{FedComp}.  Thus we are using the $v$'s to weight the points of the new configuration space.  Of the potentials in \cite{BrydgesFed} we keep only $V_r$, given in the Appendix of \cite{BrydgesFed}, in eq (A1).  It is constructed from $v_{r2}$ a two-body potential as follows
\begin{align}
v_{r2} \left( \left\{ i,j \right\} , \left\{ k,l \right\} \right) = \left\{ \begin{array}{l} 0 \quad \left\{ i,j \right\} \cap \left\{ k,l \right\} = \varnothing \\ + \infty \quad \mathrm{otherwise} \end{array} \right. .
\end{align}
Then $u \left( \left\{ i,j \right\} , \left\{ k,l \right\} \right)$ as defined in (A2) of \cite{BrydgesFed} becomes
\begin{align}
u \left( \left\{ i,j \right\} , \left\{ k,l \right\} \right) = \left\{ \begin{array}{l} 0 \quad \left\{ i,j \right\} \cap \left\{ k,l \right\} = \varnothing \\ -1 \quad \mathrm{otherwise} \end{array} \right. .
\end{align}
A natural generalization of (6) of \cite{BrydgesFed} is given by
\begin{align}
\| u \| = \sup\limits_j \left( \sum_{\left\{ k,l \right\}} \lvert v \left( k,l \right) \rvert \ \lvert u \left( \left\{ i,j \right\} , \left\{ k,l \right\} \right) \rvert \right).
\end{align}

It is easy to see from the definition of $v (i,j)$ that
\begin{align}
\| u \| \le 4
\end{align}
since
\begin{align}
\sum_j \lvert v (i,j) \rvert \le 2.
\end{align}

The generalization of (56) of \cite{BrydgesFed} easily leads to 
\begin{align}
\label{gen56}
\lvert \bar J_n \rvert \le e^n 4^n.
\end{align}

For $d=1$ the expansion in \eqref{ArrangePowerSeries} holds for all $0 \le p \le 1$, as was noted at the end of \cite{FedFried}.  We may expect this is true for all $d$!?  The methods of the current paper do not get near this result.  But the result we have encourages research to address this question.  For that matter is $\lambda_d (p)$ analytic in both $p$ and $1 / d $ for $\lvert 1 / d \rvert < 1 $, $\lvert p \rvert < 1$?  Or on the other hand perhaps the result of this paper is essentially the best one can do!

\end{document}